\documentclass[conference]{IEEEtran}
\IEEEoverridecommandlockouts
\usepackage{cite}
\usepackage{amsmath,amssymb,amsfonts}
\usepackage{algorithmic}
\usepackage{graphicx}
\usepackage{textcomp}
\usepackage{xcolor}

\usepackage{hyperref}

\usepackage{kotex}
\usepackage{multirow}
\usepackage{multicol}
\usepackage{booktabs}

\newcommand{\ie}{\textit{i}.\textit{e}.}
\newcommand{\eg}{\textit{e}.\textit{g}.}

\def\BibTeX{{\rm B\kern-.05em{\sc i\kern-.025em b}\kern-.08em
    T\kern-.1667em\lower.7ex\hbox{E}\kern-.125emX}}
\begin{document}

\title{Source-free Subject Adaptation \\ for EEG-based Visual Recognition

\thanks{
This work was supported by Institute for Information \& Communications Technology Planning \& Evaluation (IITP) grant funded by the Korea government (MSIT) (No. 2017-0-00451: Development of BCI based Brain and Cognitive Computing Technology for Recognizing Users Intentions using Deep Learning, No. 2020-0-01361: Artificial Intelligence Graduate School Program (YONSEI UNIVERSITY)).
}

}

\makeatletter
\def\footnoterule{\kern-3\p@
  \hrule \@width 2in \kern 2.6\p@} 
  
\newcommand{\linebreakand}{%
  \end{@IEEEauthorhalign}
  \hfill\mbox{}\par
  \mbox{}\hfill\begin{@IEEEauthorhalign}
}
\makeatother

\author{\IEEEauthorblockN{Pilhyeon Lee}
\IEEEauthorblockA{\textit{Department of Computer Science} \\
\textit{Yonsei University} \\
Seoul, Republic of Korea \\
lph1114@yonsei.ac.kr}
\and
\IEEEauthorblockN{Seogkyu Jeon}
\IEEEauthorblockA{\textit{Department of Computer Science} \\
\textit{Yonsei University} \\
Seoul, Republic of Korea \\
jone9312@yonsei.ac.kr}
\and
\IEEEauthorblockN{Sunhee Hwang}
\IEEEauthorblockA{\textit{Vision AI Team} \\
\textit{LG Uplus}\\
Seoul, Republic of Korea \\
sunheehwang@lguplus.co.kr}
\linebreakand
\IEEEauthorblockN{Minjung Shin}
\IEEEauthorblockA{\textit{Graduate School of Artificial Intelligence} \\
\textit{Yonsei University} \\
Seoul, Republic of Korea \\
smj139052@yonsei.ac.kr}
\and
\IEEEauthorblockN{Hyeran Byun\textsuperscript{*$\dagger$}}
\thanks{\textsuperscript{*}Corresponding author.}
\thanks{\textsuperscript{$\dagger$}Also with Graduate School of Artificial Intelligence and Graduate Program of Cognitive Science, Yonsei University.}
\IEEEauthorblockA{\textit{Department of Computer Science} \\
\textit{Yonsei University} \\
Seoul, Republic of Korea \\
hrbyun@yonsei.ac.kr}
}

\maketitle

\begin{abstract}
This paper focuses on subject adaptation for EEG-based visual recognition.
It aims at building a visual stimuli recognition system customized for the target subject whose EEG samples are limited, by transferring knowledge from abundant data of source subjects.
Existing approaches consider the scenario that samples of source subjects are accessible during training.
However, it is often infeasible and problematic to access personal biological data like EEG signals due to privacy issues.
In this paper, we introduce a novel and practical problem setup, namely source-free subject adaptation, where the source subject data are unavailable and only the pre-trained model parameters are provided for subject adaptation.
To tackle this challenging problem, we propose classifier-based data generation to simulate EEG samples from source subjects using classifier responses.
Using the generated samples and target subject data, we perform subject-independent feature learning to exploit the common knowledge shared across different subjects.
Notably, our framework is generalizable and can adopt any subject-independent learning method.
In the experiments on the EEG-ImageNet40 benchmark, our model brings consistent improvements regardless of the choice of subject-independent learning. 
Also, our method shows promising performance, recording top-1 test accuracy of 74.6\% under the 5-shot setting even without relying on source data.
Our code can be found at \href{https://github.com/DeepBCI/Deep-BCI/tree/master/1_Intelligent_BCI/Source_Free_Subject_Adaptation_for_EEG}{https://github.com/DeepBCI/Deep-BCI}.

\end{abstract}

\begin{IEEEkeywords}
Brain-computer interface, Electroencephalography, EEG-based visual recognition, Source-free subject adaptation, Deep learning
\end{IEEEkeywords}

\section{Introduction}
Brain-computer interface~(BCI) is a research topic about how to interpret and utilize brain activities of humans using computer systems and it has been actively studied due to the wide range of applications such as human intention recognition~\cite{yue2021exploring, fang2021learning, zhang2018cascade} and sleep stage classification~\cite{lee2021improving, dd_sbi, hwang2021bci, dd_2}.
Specifically, researchers have devoted a lot of efforts to decode the human mind from brain signals~\cite{tirupattur2018thoughtviz, kim2021rank}
For instance, some work allows users to control BCI-related devices like prosthetic hands by reading the users' intentions from brain signals~\cite{hosni2019eeg, park2021development}.
Meanwhile, other studies focus on extracting visual information in the brain signals and identifying the visual stimulus that users have perceived~\cite{lee2021subject-adaptive-EEG, Davis_2022_CVPR, eeg_imgnet}.

In order to develop BCI systems, several brain signal recording methods have been explored, \eg, electroencephalography~(EEG), magnetoencephalography~(MEG), and functional magnetic resonance imaging~(fMRI).
Among them, EEG is widely selected as the recording method thanks to its non-invasive property and convenience of the acquisition process, facilitating research on various fields such as emotion recognition~\cite{seed_sbd, seed}, zero-training~\cite{choi2021meta, zsl_sbi}, and motor imagery~\cite{mi1_sbd}.
This paper tackles the problem of visual stimuli recognition based on EEG signals.

\begin{figure}[t]
    \centering
    \includegraphics[clip=true, width=0.75\columnwidth]{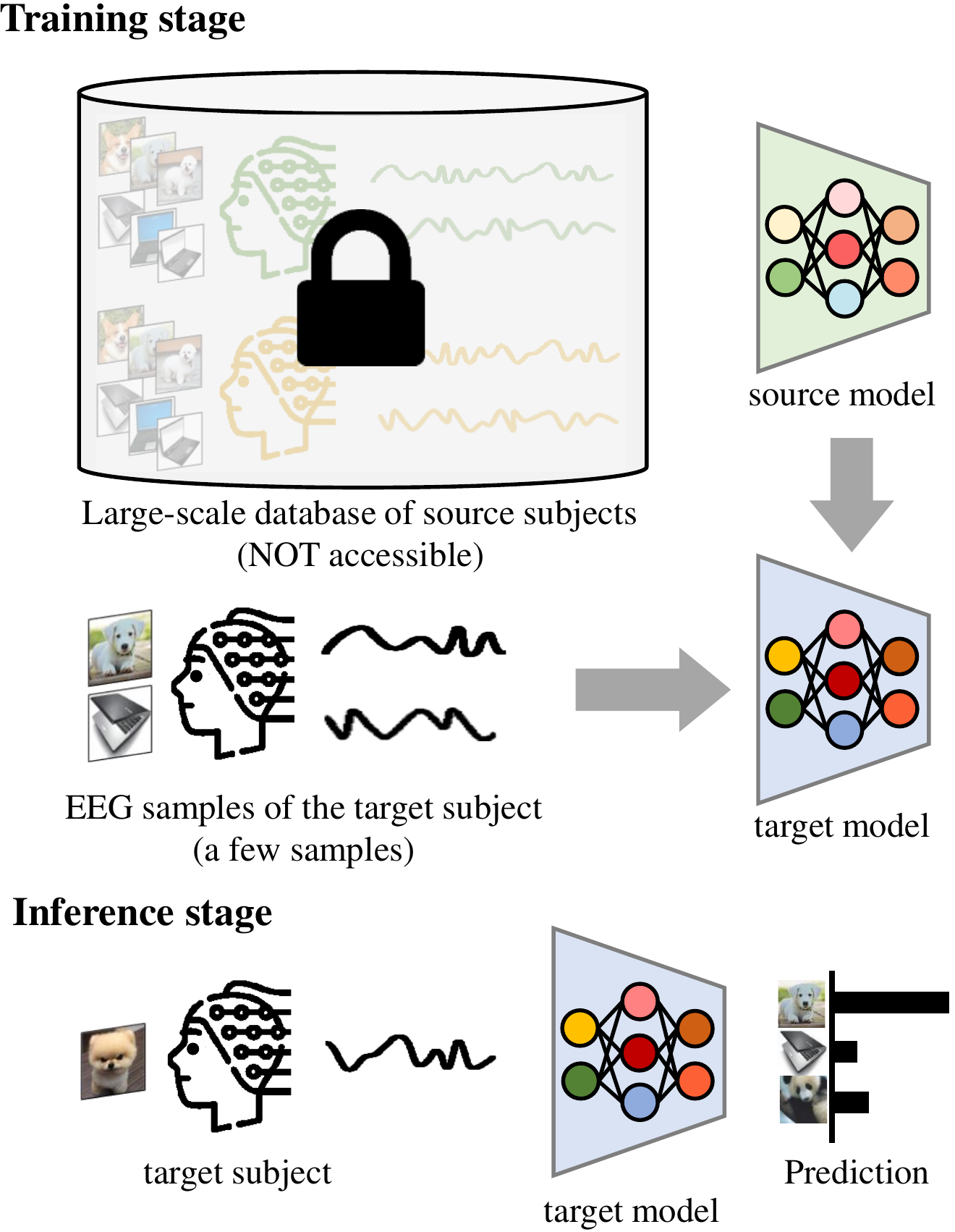}
    \caption{An illustration of the proposed source-free subject adaptation scenario. During training, we do not have access to abundant samples for source subjects. Instead, we have the pre-trained source model and only a handful of samples for the target subject. At the inference stage, the target model needs to accurately predict the category of the visual stimuli based on EEG signals from the target subject.}
    \label{fig:intro_fig}
    \vspace{-3mm}
\end{figure}

In the last years, with the success of deep learning~\cite{he2016deep,simonyan2014vgg16}, EEG-based visual stimuli recognition systems have shown remarkable performance improvements~\cite{Davis_2022_CVPR, ahn2022decoding, seed_sbd, dd_sbi}.
However, they are generally dependent on subjects; that is, they perform well only for the existing users in the training database but show poor performance for new users.
To handle this issue, the subject adaptation paradigm has recently been proposed to alleviate the cost of calibration process~\cite{lee2021subject-adaptive-EEG, lee2022inter}, where only a handful of EEG samples are acquired from the target subject while abundant data from other source subjects are available.
With sophisticated subject-independent feature learning, recent methods achieve promising performance~\cite{lee2021subject-adaptive-EEG, lee2022inter}, demonstrating the potential for efficient adaptation of EEG-based visual recognition models to new users.

Despite the great performance, existing subject adaptive visual recognition models hinge on the impractical assumption that source subject data are available during training.
In many real-world scenarios, however, personal medical data such as EEG signals contain private information, and thus transmitting and sharing them violates the data privacy policy.
This restricts the existing methods from being deployed in real-world systems.
To mitigate the restriction, we introduce a novel yet practical problem setting of EEG-based visual recognition, namely \textit{source-free subject adaptation}.
In specific, EEG data from source subjects are inaccessible and only pre-trained model weights can be utilized for model adaptation.
An illustration of the proposed setting is provided in Fig.~\ref{fig:intro_fig}.

It is challenging to build an adapted visual recognition model for the target subject without accessing the large-scale source dataset.
To bypass the reliance on the source subject data, we propose a generative method based on the pre-trained source model weights.
Specifically, we train a generator to synthesize EEG samples by maximizing their predicted likelihood from the classifier of the source model.
After training, the generator is capable of producing synthetic data that simulate the real samples from source subjects and it can control the class labels.
Then the generated samples serve as pseudo-source data to be utilized for subject-independent representation learning along with the handful of target subject data.
Our framework is general and therefore can adopt any subject-independent training method; we use two representative methods~\cite{lee2021subject-adaptive-EEG,lee2022inter} for the experiments.
In the experimental results on EEG-ImageNet40~\cite{eeg_imgnet}, we verify the effectiveness of the proposed framework under the source-free subject adaptation scenario.
Notably, our method brings significant performance improvements over the baseline even without access to source data under various settings.
In addition, our model even performs comparably or even favorably compared to the conventional subject adaptation methods that rely on the source data.

We summarize our contributions as follows.
\begin{itemize}
    \item We propose a new practical setting of subject adaptive visual recognition, where source subject data are inaccessible during adaptation.
    \item We design a generative approach to perform subject adaptation under the source-free scenario. Our method is generalizable to feature learning methods and shows consistent improvements over the baseline.
    \item Experiments under various settings showcase the efficacy of the proposed method. It records the top-1 accuracy of 74.6\% in the 5-shot setting on EEG-ImageNet40.
\end{itemize}

\section{Related Work}
\subsection{Brain activities with visual perceptions}
Recognizing visual perception from brain activities aims to decode visual working memory.
Since the visual cortex occupies a large portion of the brain, brain signals carry a vast amount of information such as objects, scenery, and characters.
In recent years, a number of studies~\cite{Ahmed_2021_CVPR, ahn2022decoding, eeg_imgnet, kim2021rank, kim2021intuitive} have been conducted to recognize the human imagination based on EEG data induced by visual stimuli.
For instance, some work tries to directly decode the own memory of humans~\cite{Ahmed_2021_CVPR, ahn2022decoding, eeg_imgnet}, while others focus on controlling devices by recognizing the visual stimulus~\cite{kim2021intuitive}.
On the other hand, with the recent advance of image generation models, recent research attempts to reconstruct visual stimuli from the brain signals~\cite{thoughtviz,brain2image, movie_recon} and enables the users to readily edit the reconstructed images by thinking~\cite{Davis_2022_CVPR}.

\subsection{Subject-adaptive EEG-based classification}
To alleviate the performance drop of the model when confronting EEG signals from new users who are not registered in the training database, previous studies~\cite{feature_sbi,feature2_sbi} focus on extracting subject-independent features from the source EEG signals. Several approaches~\cite{em_sbi,sd_sbi,dd_sbi} hinge on adversarial training-based strategies to encourage the subject invariance as well as the class discriminability of the feature representation during the training process.
Meanwhile, there appear the studies on subject adaptation~\cite{lee2021subject-adaptive-EEG, lee2022inter}, aiming at correctly recognizing the target class only with a few EEG signals from the new users using a sufficient number of pre-collected data from source subjects. To this end, they exploit domain adaptation methods~\cite{li2019domain, da_aaai} and contrastive learning~\cite{chen2020simple,khosla2020supcon} to narrow the representation gap between the source subjects and the target one.
In contrast to them, we point out that source EEG data have a large chance to be unavailable for model training due to privacy issues.
To handle this challenge, we propose a generative approach to synthesize fake EEG samples that can replace the real ones during the subject adaptation process.

\section{Method}
\label{sec:method}

We start by formulating the proposed source-free problem setting of the subject adaptive EEG-based visual recognition.
Its goal is exactly the same with the conventional setting: we want to effectively adapt a model to the target subject when there are only a limited amount of data $\mathcal{D}^{trg}$ from the target subject.
Meanwhile, the major difference lies in that pre-trained model weights on the source subject dataset $\mathcal{D}^{src}$ instead of raw data are available during training.
Therefore, we need to build an adapted visual recognition model $M^{trg}$ for the target subject based on the pre-trained source model $M^{src}$ and the small target dataset $\mathcal{D}^{trg}$, while $\mathcal{D}^{src}$ is withheld.
In the following, we elaborate on the training of the source model $M^{trg}$.
Thereafter, we introduce the proposed source data generation and the target subject adaptation.

\subsection{Source model training}
Our EEG-based visual recognition model $M^{src}$ consists of a sequence encoder $f$, an embedding layer $g$, and a classifier $h$. Concretely, the sequence encoder $f$, composed of a single gated recurrent unit~(GRU), encodes the input temporal EEG signal $x \in \mathbb{R}^{T \times D_{in}}$ into the feature representation, \ie, $v=f(x)\in \mathbb{R}^{D_{enc}}$, where $T$ is the length of the EEG sample, while $D_{in}$ and ${D}_{enc}$ indicate the input and the output dimensions, respectively. Note that we use only the output feature from the GRU unit at the final timestamp. Afterward, the embedding layer $g$ maps the representation $v_i$ into the embedding space, resulting in the embedded feature $w=g(z) \in \mathbb{R}^{D_{emb}}$, where $D_{emb}$ denotes the embedding dimension. In the end, the classifier $h$ takes $w$ as input and predicts the class probability $p(\mathbf{y}|x;\theta^{src})=h(w) \in \mathbb{R}^{C}$, where $\theta^{src}$ is all the trainable parameters in the model $M^{src}$ and $C$ indicates the total number of classes.

Based on the assumption of the source-free setting, the source model is first trained on the plentiful data from source subjects using the classification loss as follows.
\begin{equation}
    \mathcal{L}_{src} = - \frac{1}{|\mathcal{D}^{src}|}\sum_{\forall (x_i, y_i) \in \mathcal{D}^{src}} y_i \log p(y_i|x_i;\theta^{src}),
    \label{equ:source}
\end{equation}
where $y_i$ is the ground-truth label of the $i$-th sample. We train the source model sufficiently until the convergence so that the model possesses strong class discriminability on signals from the source subjects.
The source model training is illustrated in Fig.~\ref{fig:contrast_fig}a.
Hereafter, the source data are no longer accessible, and only the trained source model $M^{src}$ and the target data $\mathcal{D}^{trg}$ containing only a few target samples ($k$-shot) are utilizable during training.

\subsection{Source data generation}
\label{sec:generator}
Directly training the target classification model with the target samples results in unsatisfactory prediction accuracy due to data scarcity.
Even if we initialize the target model using the pre-trained source model weights, the performance is not likely to be improved significantly.
This is because of the feature discrepancy between the source and the target subjects.
Directly narrowing the feature gap may be a promising direction, but it is nontrivial in the case that the source subject data are inaccessible.

To this end, we train a generator $G$ which can synthesize the source data to implicitly approximate the source data distribution. Specifically, we first sample a latent code $z \in \mathbb{R}^{D_z}$ from the latent space, \ie, $z\sim Z$, where $D_z$ is the dimension of the latent space. We feed the latent code $z$ and the randomly sampled one-hot class code $c \in \mathbb{R}^{C}$ to the generator and obtain the generator output $G(z, c) \in \mathbb{R}^{T \times D_{in}}$. Afterward, we utilize the classifier responses of the pre-trained source model $M^{src}$ to train the generator to produce synthetic source data.
The loss function of the generator $G$ is formulated by the cross-entropy loss as follows.
\begin{equation}
    \mathcal{L}_{G} = -  \sum_{\forall i} c_i \log p(c_i|G(z_i, c_i);\theta^{src}).
    \label{equ:generator}
\end{equation}
Here $p(c_i|G(z_i, c_i);\theta^{src})$ denotes the output prediction of the frozen source model with its parameters $\theta^{src}$ given the generated sample by $G$. Notably, the frozen source model $M^{src}$ can be viewed as a kind of discriminator that judges whether the generated sample corresponds to the input class code $c$ in order to guide the generator $G$.
Therefore, the generator $G$ is encouraged to produce generated samples that follow the source distribution in order to minimize the loss.
The training of the generator is depicted in Fig.~\ref{fig:contrast_fig}b.
After training, we can utilize the generated samples as pseudo-source samples for subject adaptation.

\begin{figure}[t]
    \centering
    \includegraphics[clip=true, width=1\columnwidth]{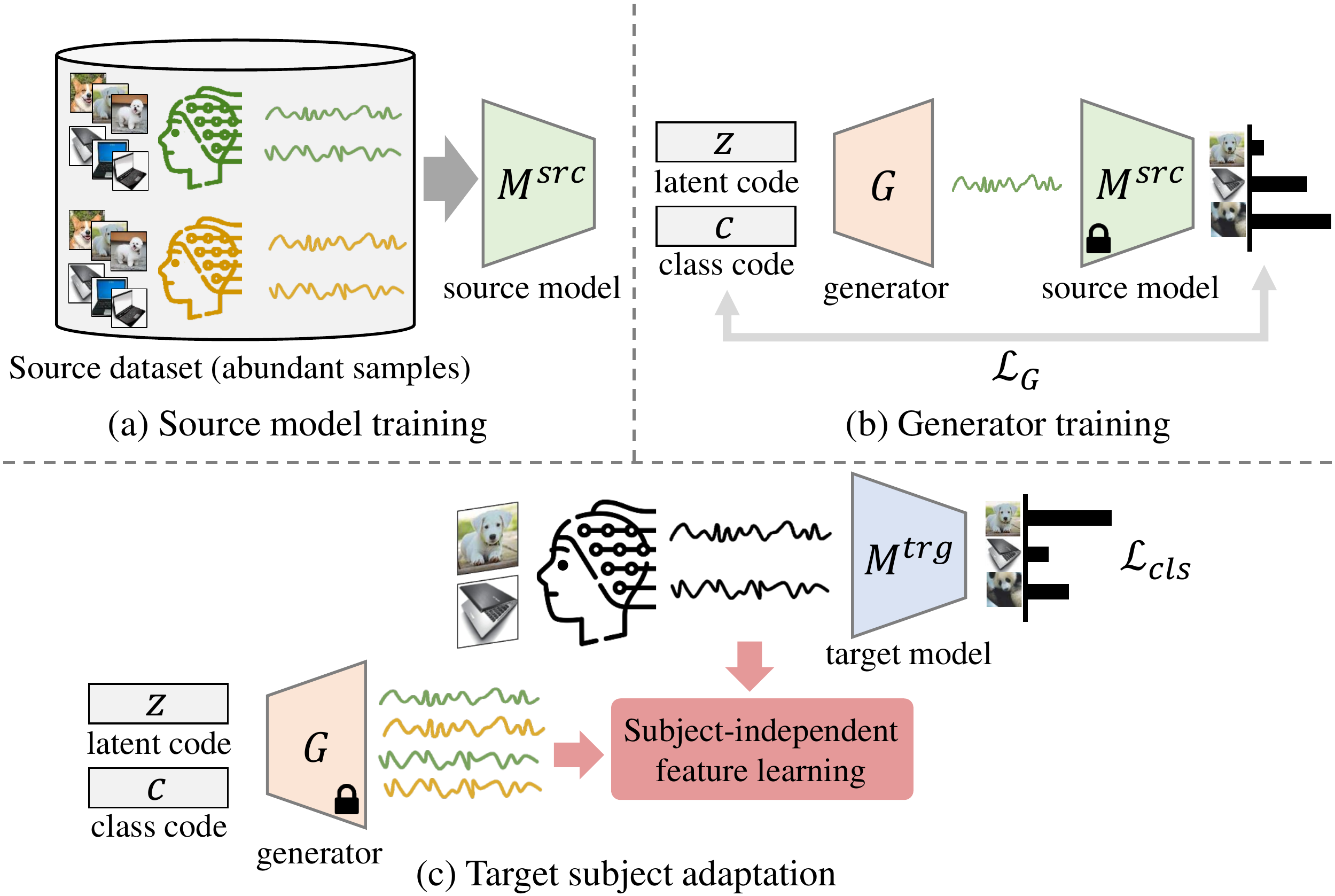}
    \caption{Illustration of our subject adaptation process under the source-free scenario.}
    \label{fig:contrast_fig}
    \vspace{-1mm}
\end{figure}

\subsection{Target subject adaptation}
Based on the synthesized fake source samples, we align the feature distribution of target samples with that of source ones in order to narrow their representation gap.
For this purpose, a variety of subject-independent training methods can be utilized.
Here we adopt two recent approaches, namely MMD-based~\cite{lee2021subject-adaptive-EEG} and contrastive learning-based subject adaptation methods~\cite{lee2022inter}.
The overview of the target subject adaptation process of the target model $M^{trg}$ is shown in Fig.~\ref{fig:contrast_fig}c.
In the following, we briefly review the subject-independent training.

\noindent\textbf{MMD-based method.}~
Maximum mean discrepancy (MMD) \cite{long2015mmd} estimates the distance (discrepancy) between two different feature distributions. Based on the recent work~\cite{lee2021subject-adaptive-EEG}, we exploit this metric to compute the gap between features from fake source subjects and the target subject.
Specifically, we first build two separate training batches, \ie, $\mathcal{B}^{src}$ and $\mathcal{B}^{trg}$, which respectively consist of the fake source samples $G(z)$ and the samples from the target subject $\mathcal{D}^{trg}$.
We calculate MMD and minimize it to align the distributions as follows.
\begin{equation}
    \mathcal{L}_\text{MMD} = \left \|\frac{1}{|\mathcal{B}^{src}|}\sum^{\mathcal{B}^{src}}_{i=1}\phi(w_i^{src}) - \frac{1}{|\mathcal{B}^{trg}|}\sum^{\mathcal{B}^{trg}}_{i=1}\phi(w_i^{trg})\right \|_{F},
    \label{equ:mmd}
\end{equation}
where $\phi(\cdot):\mathcal{W}\rightarrow\mathcal{F}$ is an embedding function to map the features into the reproducing kernel Hilbert space (RKHS), and $\left\|\cdot\right\|_{F}$ is the Frobenius norm.
$w_i^{src}$ and $w_i^{trg}$ indicate the feature representations of the $i$-th samples that are drawn from $\mathcal{B}^{src}$ and $\mathcal{B}^{trg}$, respectively.

\noindent\textbf{Contrastive learning-based method.}~
Contrastive learning is originally designed to explicitly model inter-sample relationships by pulling or pushing sample representations in the feature space~\cite{chen2020simple,khosla2020supcon}.
Thanks to its effectiveness, it has been widely used in various research fields~\cite{wu2021contrastive,lee2021completeness}.
In this work, we adopt inter-subject contrastive learning (IS-Con)~\cite{lee2022inter} to maximize the similarity between the features from the same class regardless of the subjects they are obtained from.
In specific, we first construct a training batch $B$ with the target samples $\mathcal{D}^{trg}$ as well as the synthesized source samples $G(z)$.
Then we optimize the target model $M^{trg}$ with the inter-subject contrastive loss, which is defined as follows.
\begin{equation}
    \mathcal{L}_{\text{con}} = - \frac{1}{|\mathcal{B}|}\sum_{\forall i \in \mathcal{B}}
    \log
    \frac{\sum_{\forall j \in \mathcal{P}(i)}
    \exp(w_i \cdot w_j / \tau)}
    {\sum_{\forall k \in \mathcal{A}(i)}
    \exp(w_i \cdot w_k / \tau)},
    \label{equ:contrastive}
\end{equation}
where $w_i$ is the embedded feature representation of the $i$-th sample, $\mathcal{A}(i)$ is the anchor set of the $i$-th sample, $\mathcal{P}(i)$ denotes the positive set that consists of the indices of samples from the same class of the anchor $x_i$, and $\tau$ is a temperature parameter that controls the sharpness of the similarity distribution.
the anchor set is defined as $\mathcal{A}(i) = \mathcal{P}(i) \cup S(i)$, where $S(i)$ is a set that contains the indices of the samples from the same subject of the anchor. This ensures that the complementary set, \ie, $\mathcal{A}(i)^c=\mathcal{P}(i)^{c} \cap \mathcal{S}(i)^{c}$, is not considered for loss computation.
According to the literature~\cite{jeon2021stylizationDG, park2022fair-contrastive}, compelling the anchor to expel the complementary set $\mathcal{A}(i)^c$ causes the representations from different subjects to be distinguishable, hence obstructing subject-independence.
Therefore, by excluding them in contrastive learning, the model could learn more subject-independent representations~\cite{lee2022inter}.

In addition to the loss for subject-independent learning above, we utilize the cross-entropy loss to enhance the class discriminability of the target samples.
\begin{equation}
    \mathcal{L}_{\text{cls}} = -  \frac{1}{|\mathcal{D}^{trg}|}\sum_{\forall (x_i, y_i) \in \mathcal{D}^{trg}} y_i \log p(y_i|x_i;\theta^{trg}),
    \label{equ:cross_entropy}
\end{equation}
where $y_i$ is the label of the sample $x_i$.

The overall loss for the target EEG classification model $M^{trg}$ is a weighted sum of two losses as follows.
\begin{equation}
    \mathcal{L}_{\text{total}} = \mathcal{L}_{\text{cls}} + \lambda\mathcal{L}_{*},
\end{equation}
where $* \in \{\text{MMD}, \text{con}\}$ indicates the choice of the subject-independent feature learning method, while $\lambda$ is a balancing factor that is empirically set to 1.

\section{Experiments}
\label{sec:experiments}

\subsection{Dataset}
In the experiments, we use EEG-ImageNet40~\cite{eeg_imgnet} as the benchmark dataset which is composed of EEG signals with the corresponding class labels of the visual stimuli. Each EEG signal has a channel size of 128 ($D_{in}=128$) and a temporal length of 480~ms with a unit step of 1~ms. For pre-processing, a band-pass filter (14-72 Hz) is applied to sieve Beta and Gamma frequency bands that are known to be related to visual perceptions, and a notch filter (49-51Hz) is utilized to get rid of noisy signals. The dataset is collected from six individual subjects (\ie, five males and one female), containing a total of 40 visual stimuli categories that are from ImageNet classes~\cite{ImageNet}. To simulate the subject adaptation scenario, we fix one subject as the target subject and set the others to be the source subjects.
The average number of samples per class is roughly 50, resulting in a total of 2,000 samples.
In the adaptation scenario, we assume that only a few samples exist for the target subject, \ie, $k$ samples for each class ($k$-shot).
In addition, our source-free setting further prohibits using the abundant source data and instead allows for utilizing the pre-trained model weights.
During the evaluation, we split the dataset into training, validation, and test subjects in accordance with the official split (4:1:1).

\subsection{Implementation details}
Following the original dataset and previous studies~\cite{eeg_imgnet, lee2021subject-adaptive-EEG, lee2022inter}, we crop and use only the interval of 320 ms to 480 ms of the raw EEG signals. Accordingly, each EEG signal has the temporal length $T$ of $160$ and the channel dimension $D_{in}$ is set to 128 as aforementioned.
The target visual recognition model $M^{trg}$ has the same architecture as the source model $M^{src}$. The generator $G$ is a multi-layer perceptron where its input, hidden, output dimensions are set to 100, 128, and 128, respectively. Note that we generate source features instead of raw EEG signals for efficiency.
In the first stage, we train the source EEG classification model $M^{src}$ with a batch size of 1,200 during 200 epochs. After that, the generator $G$ is optimized with a batch size of 1,200 during 200 epochs. Finally, the target model $M^{trg}$ is trained in the same way with $M^{src}$ for 200 epochs. For optimization, the Adam optimizer~\cite{kingma2014adam} is utilized with a learning rate of 0.001.

We implement the MMD~\cite{long2015mmd} and IS-Con~\cite{lee2022inter} losses following their official codebase. In the remaining section, we experiment with two variants of our framework that respectively use two different target subject adaptation techniques: $\text{Ours}_{\text{MMD}}$ and $\text{Ours}_{\text{con}}$.

\subsection{Quantitative results}

\subsubsection{Comparison in the 1-shot setting}
In Table~\ref{table:quant_subject}, we compare the performance of our method on different target subjects to verify the effectiveness in the challenging 1-shot setting.
The entries are grouped based on whether they utilize source data or not.
Although it is unfair to compare, we present the results of conventional models~\cite{lee2021subject-adaptive-EEG,lee2022inter} that rely on source data for reference.
The first thing we observe from the results is that the $k$-shot baseline shows deprived performance on both validation and test sets due to the harsh constraints of the source-free adaptation scenario (\ie, w/o source data).
In contrast, both the proposed MMD-based and contrastive learning-based models, \ie, $\text{Ours}_{\text{MMD}}$ and $\text{Ours}_{\text{con}}$, consistently improve the performance for every target subject.
Meanwhile, it is observed that the overall accuracy gain from $\text{Ours}_{\text{MMD}}$ is smaller than that from $\text{Ours}_{\text{con}}$.
On the other hand, $\text{Ours}_{\text{con}}$ successfully adapts the model to the target subject by transferring the knowledge from synthesized source samples.
Moreover, our model sometimes performs comparably or even favorably against the conventional MMD-based method~\cite{lee2021subject-adaptive-EEG} that rely on the large-scale source dataset, manifesting the effectiveness of our generative approach in the source-free scenario.


\begin{table}[t]
\caption{
Top-1 accuracy comparison by changing the target subject in the 1-shot setting. Entries are categorized into two groups with and without source data access.
}
\centering
\resizebox{1.0\columnwidth}{!}{
\begin{tabular}{c|ccc|ccc}
\toprule
\multicolumn{7}{c}{Validation set} \\ \midrule
\multirow{2}{*}{Subject}  & \multicolumn{3}{c|}{w/o source data} & \multicolumn{3}{c}{w/ source data} \\ 
       & $k$-shot  & Ours$_{\text{MMD}}$     & Ours$_{\text{con}}$   & Vanilla     & MMD~\cite{lee2021subject-adaptive-EEG}   & IS-Con~\cite{lee2022inter} \\ \midrule
\#0 & $13.5_{\pm2.1}$   & $17.7_{\pm2.8}$   & $\textbf{31.2}_{\pm1.9}$   & $29.3_{\pm1.9}$   & $35.7_{\pm1.9}$    & $39.2_{\pm1.8}$   \\
\#1 & $12.6_{\pm2.1}$   & $14.7_{\pm2.8}$   & $\textbf{29.6}_{\pm3.1}$   & $21.8_{\pm2.3}$   & $29.0_{\pm3.6}$    & $33.0_{\pm2.6}$   \\
\#2 & $17.0_{\pm1.6}$   & $20.9_{\pm2.6}$   & $\textbf{31.0}_{\pm1.4}$   & $25.3_{\pm0.9}$   & $30.8_{\pm2.2}$    & $36.9_{\pm2.2}$   \\
\#3 & $27.8_{\pm1.7}$   & $31.2_{\pm1.8}$   & $\textbf{36.8}_{\pm2.3}$   & $28.8_{\pm2.2}$   & $31.9_{\pm3.9}$    & $43.6_{\pm4.7}$   \\
\#4 & $16.3_{\pm2.8}$   & $22.3_{\pm3.2}$   & $\textbf{36.7}_{\pm1.9}$   & $25.9_{\pm1.9}$   & $36.2_{\pm3.3}$    & $40.0_{\pm3.3}$   \\
\#5 & $9.2_{\pm1.4}$   & $12.1_{\pm2.1}$   & $\textbf{24.4}_{\pm2.4}$   & $20.7_{\pm2.9}$   & $25.8_{\pm1.7}$    & $34.8_{\pm1.9}$  \\ \midrule

\multicolumn{7}{c}{Test set} \\ \midrule
\multirow{2}{*}{Subject}  & \multicolumn{3}{c|}{w/o source data} & \multicolumn{3}{c}{w/ source data} \\ 
       & $k$-shot  & Ours$_{\text{MMD}}$     & Ours$_{\text{con}}$   & Vanilla     & MMD~\cite{lee2021subject-adaptive-EEG}   & IS-Con~\cite{lee2022inter} \\ \midrule
\#0 & $12.2_{\pm2.1}$   & $14.9_{\pm2.7}$   & $\textbf{28.5}_{\pm0.6}$   & $24.3_{\pm0.9}$   & $29.6_{\pm4.9}$    & $32.8_{\pm4.5}$      \\
\#1 & $10.3_{\pm2.2}$   & $13.1_{\pm2.9}$   & $\textbf{26.9}_{\pm3.9}$    & $18.1_{\pm2.7}$   & $25.4_{\pm2.4}$    & $29.2_{\pm3.1}$   \\
\#2 & $15.5_{\pm2.9}$   & $17.6_{\pm1.9}$   & $\textbf{27.8}_{\pm2.2}$    & $23.9_{\pm3.0}$   & $29.2_{\pm3.7}$    & $35.6_{\pm3.8}$   \\
\#3 & $26.2_{\pm3.2}$   & $29.0_{\pm2.8}$   & $\textbf{34.1}_{\pm1.6}$    & $27.4_{\pm3.2}$   & $32.1_{\pm4.3}$    & $40.4_{\pm4.2}$   \\
\#4 & $15.2_{\pm1.9}$   & $20.0_{\pm2.3}$   & $\textbf{34.1}_{\pm2.5}$    & $22.7_{\pm1.2}$   & $35.3_{\pm3.6}$    & $37.5_{\pm3.7}$   \\
\#5 & $7.0_{\pm1.0}$   & $9.4_{\pm1.0}$   & $\textbf{20.6}_{\pm2.0}$    & $18.9_{\pm2.9}$   & $21.4_{\pm2.6}$    & $30.2_{\pm2.9}$   \\
    \bottomrule
\end{tabular}
}
\label{table:quant_subject}
\end{table}

\subsubsection{Comparison with varying k}
In Table~\ref{table:quant_k}, we quantitatively compare the performance for the target subject with varying $k$. Noticeably, the performance gain from our framework is consistent regardless of the number of target samples $k$ during training. Especially, delightful accuracy improvements are brought by both $\text{Ours}_{\text{MMD}}$ and $\text{Ours}_{\text{con}}$ especially in low-shot settings (\eg, $k=1,2,3$), emphasizing the efficacy of the proposed framework.
On the other hand, $\text{Ours}_{\text{MMD}}$ shows poor performance when $k$ is large.
We suspect that the generated samples have insufficient diversity to represent the whole source data distribution, leading to limited benefits from the inter-subject feature learning as $k$ becomes large.
On the contrary, $\text{Ours}_{\text{con}}$ brings consistent and substantial improvements across all $k$, even sometimes outperforming the conventional methods~\cite{lee2021subject-adaptive-EEG,lee2022inter} that use raw source data.
This indicates that inter-subject contrastive learning is highly effective in our generative framework, as it ensures class discriminability and subject independence of the features and thus allows for building the target-adapted model in a data-efficient manner.

\begin{table}[t]
\caption{
Top-1 accuracy comparison by changing the amount of target data ($k$-shot). Entries are categorized into two groups with and without source data access.
}
\centering
\resizebox{0.93\columnwidth}{!}{
\begin{tabular}{c|ccc|ccc}
\toprule
\multicolumn{7}{c}{Validation set} \\ \midrule
\multirow{2}{*}{$k$}  & \multicolumn{3}{c|}{w/o source data} & \multicolumn{3}{c}{w/ source data} \\ 
       & $k$-shot     & Ours$_{\text{MMD}}$     & Ours$_{\text{con}}$      & Vanilla     & MMD~\cite{lee2021subject-adaptive-EEG}      & IS-Con~\cite{lee2022inter}  \\ \midrule
1 & $16.0_{\pm0.6}$   & $19.8_{\pm1.2}$   & $\textbf{31.6}_{\pm0.6}$    & $25.3_{\pm1.0}$   & $31.7_{\pm1.5}$    & $37.9_{\pm3.5}$   \\
2 & $33.2_{\pm1.2}$   & $35.4_{\pm0.6}$   & $\textbf{48.7}_{\pm0.9}$    & $41.7_{\pm1.9}$   & $46.3_{\pm1.8}$    & $54.3_{\pm4.5}$   \\
3 & $49.9_{\pm0.4}$   & $48.4_{\pm0.8}$   & $\textbf{61.5}_{\pm1.6}$    & $54.4_{\pm1.0}$   & $58.9_{\pm0.7}$    & $64.7_{\pm4.4}$   \\
4 & $61.9_{\pm2.0}$   & $58.2_{\pm1.2}$   & $\textbf{70.0}_{\pm1.7}$    & $64.6_{\pm1.5}$   & $67.5_{\pm1.2}$    & $71.6_{\pm4.9}$   \\
5 & $70.0_{\pm1.6}$   & $64.7_{\pm1.5}$   & $\textbf{76.1}_{\pm1.4}$    & $72.0_{\pm1.3}$   & $73.5_{\pm1.1}$    & $76.1_{\pm4.2}$   \\ \midrule
\multicolumn{6}{c}{Test set} \\ \midrule
\multirow{2}{*}{$k$}  & \multicolumn{3}{c|}{w/o source data} & \multicolumn{3}{c}{w/ source data} \\ 
       & $k$-shot     & Ours$_{\text{MMD}}$     & Ours$_{\text{con}}$      & Vanilla     & MMD~\cite{lee2021subject-adaptive-EEG}      & IS-Con~\cite{lee2022inter}  \\ \midrule
1 & $14.4_{\pm1.6}$   & $17.3_{\pm1.4}$   & $\textbf{28.5}_{\pm0.6}$    & $22.5_{\pm0.8}$   & $28.8_{\pm1.2}$    & $34.3_{\pm4.0}$   \\
2 & $31.2_{\pm1.2}$   & $33.3_{\pm1.3}$   & $\textbf{47.4}_{\pm2.2}$    & $39.9_{\pm2.0}$   & $43.8_{\pm1.4}$    & $51.2_{\pm5.0}$      \\
3 & $48.2_{\pm2.6}$   & $46.2_{\pm2.1}$   & $\textbf{60.2}_{\pm2.0}$    & $52.6_{\pm1.7}$   & $56.4_{\pm1.7}$    & $62.2_{\pm5.2}$      \\
4 & $60.4_{\pm0.9}$   & $56.2_{\pm1.1}$   & $\textbf{68.6}_{\pm1.9}$    & $62.4_{\pm1.7}$   & $64.7_{\pm1.6}$    & $68.6_{\pm5.1}$      \\
5 & $68.1_{\pm1.6}$   & $62.7_{\pm1.8}$   & $\textbf{74.6}_{\pm2.1}$    & $69.5_{\pm1.1}$   & $70.1_{\pm1.0}$    & $72.6_{\pm4.2}$      \\
    \bottomrule
\end{tabular}

}
\label{table:quant_k}
\end{table}

\section{Conclusion}
In this paper, we pointed out the impractical problem setting of the existing subject adaptation for EEG-based visual recognition approaches.
Then we introduce a novel yet practical setting, where source data are not allowed for subject adaptation. To accomplish source-target subject alignment without any source data, we proposed a generative framework that synthesizes fake source data based on the pre-trained source model weights. Afterwards, we apply subject adaptation techniques, \ie, MMD minimization and contrastive learning, to achieve learning subject-invariant representation.
As a result, the proposed method brings large performance improvements over the baseline on the EEG-ImageNet40 dataset.
Moreover, it sometimes performs comparably compared to the existing subject-adaptive visual recognition models that rely on a large amount of source data.

\bibliographystyle{IEEEtran}
\bibliography{ref}

\end{document}